\documentclass[letterpaper,twocolumn,english]{revtex4-1}
\pdfoutput=1
\usepackage[T1]{fontenc}
\usepackage[latin9]{inputenc}
\usepackage{xcolor}
\usepackage{pdfcolmk}
\usepackage{booktabs}
\usepackage{bm}
\usepackage{amsmath}
\usepackage{amssymb}
\usepackage{graphicx}
\PassOptionsToPackage{normalem}{ulem}
\usepackage{ulem}
\usepackage{subscript}

\makeatletter

\begin{document}

\title{Microscopic theory of  the polarizability of transition metal dichalcogenides excitons: Application to WSe\textsubscript{2}}

\author{J. C. G. Henriques$^{1,2}$, M. F. C. Martins Quintela$^{1}$, N.
M. R. Peres$^{1,2*}$}
\address{$^{1}$Department and Centre of Physics, and QuantaLab, University
of Minho, Campus of Gualtar, 4710-057, Braga, Portugal}
\address{$^{2}$International Iberian Nanotechnology Laboratory (INL), Av. Mestre
José Veiga, 4715-330, Braga, Portugal}

\begin{abstract}
In this paper we develop a fully microscopic theory of the polarizability of excitons in transition metal dichalcogenides.  We apply our method to the description of the excitation  $2$p  dark states. These states are not observable in absorption experiments but can be excited in a pump-probe experiment.  As an example we consider  $2$p  dark states in WSe\textsubscript{2}.
We find a good agreement between recent experimental measurements and our theoretical calculations.
\end{abstract}

\maketitle
\section{\label{sec: Intro}Introduction}

The optical properties of monolayer transition metal dichalcogenides
(TMDs) are of considerable interest \emph{per se}, but also from the
point of view of applications \cite{lv_transition_2015,ma_tunable_2020,mueller_exciton_2018,wang_colloquium_2018,mak_lightvalley_2018,schneider_two-dimensional_2018}.
It is by now well known that the optically bright exciton absorption
peaks correspond to the excitation of states in the $n$s series \cite{schneider_two-dimensional_2018,hsu_dielectric_2019},
with the 1s being energy split due to the strong spin-orbit effect
present in these systems. The excitons in the $n$p series are optically
dark in TMDs and a very faint d-exciton line has also been predicted
\cite{Chaves_2017} but never observed to date. The dark excitons
can, however, be controlled magnetically \cite{zhang_magnetic_2017}.
The Berry phase-induced splitting of the 2p states in MoSe\textsubscript{2}
was studied in Ref. \cite{yong_valley-dependent_2019}.

The dielectric response of TMDs has two different regimes. The first,
named the \emph{interband}, occurs when an electron in the valence
band is promoted to the conduction band. Due to the attractive electrostatic
interaction between the hole left in the valence band and the electron
promoted to the conduction band, this excitation leads to the formation
of excitonic states \cite{steinleitner_direct_2017,hsu_dielectric_2019}.
From those, the $ns$ are optically active and can be seen in absorption
measurements \cite{Potemski2017}. The other possible regime, which
we call the \emph{intra-exciton transitions}, consists of transitions
between the occupied $1s$ state and the empty $np$ states of the
excitonic energy levels \cite{miyajima_optical_2016,berghauser_optical_2016}.
Each of the $1s\rightarrow np$ transitions \cite{berghauser_mapping_2018,yong_valley-dependent_2019,hsu_dielectric_2019,merkl_ultrafast_2019}
is characterized by an in-plane polarizability (a quantity shown relevant
in other semiconductors \cite{tian_electronic_2020,wang_exciton_2006}),
which in turn determines the dielectric response of the system in
a pump-probe experiment, using lasers of different frequencies. In
this work, we consider the case of the TMD WSe\textsubscript{2} with
which experiments similar to the ones just described have been performed
\cite{poellmann_resonant_2015}. In Ref. \cite{cha_1_2016} MoS\textsubscript{2}
was studied. Also, valley-selective physics has been measured in this
system \cite{sie_valley-selective_2015,mak_lightvalley_2018}. In
Ref. \cite{poellmann_resonant_2015}, a 90-fs laser pulse centered
at $\lambda_{c}=742$ nm was shined on a WSe\textsubscript{2}monolayer
leading to the population of the $1s$ excitonic state. At a certain
variable time delay, the low-energy dielectric response is probed
by a phase-locked mid-infrared pulse.

The first experimental evidence of $1s\rightarrow np$ transitions
between exciton levels of the associated Rydberg series \cite{frohlich_observation_1985}
was found in the solid state system Cu\textsubscript{2}O, at the
temperature of 1.8 K \cite{frohlich_observation_1985}. More recently,
the experiment of Poellmann \emph{et al.} \cite{poellmann_resonant_2015}
showed that the same physics can be observed at room temperature in
WSe\textsubscript{2}, a work that sprung interest by other groups
\cite{zhang_experimental_2015,merkl_ultrafast_2019}. The results
of Poellmann \emph{et al.} are special due to the two-dimensional
nature of this TMD, which entails a poorer electrostatic screening
between the electron and the hole brought about by the reduced dimension
of WSe\textsubscript{2} when compared to the three dimensional case
of Cu\textsubscript{2}O.

The field-resolved time-domain data allowed the retrieving of the
full dielectric response of the excited sample \cite{poellmann_resonant_2015}.
The dielectric response is characterized by by the real and imaginary
parts of dielectric function $\Delta\epsilon(\omega)=\epsilon_{1}(\omega)+i\epsilon_{2}(\omega)$.
The real part follows a dispersive shape with a zero crossing at a
photon energy of 180 meV, for a given exciton density (laser fluence).
As the authors stress, this spectrum is in striking contrast to the
Drude-like response of free electron\textendash hole pairs observed
after non-resonant above-band gap excitation, and is, therefore, a
clear signature of the resonant excitation of intra-exciton transition
$1$s$\rightarrow 2$p referred above. The authors have also mapped
the real part of the optical conductivity, $\Delta\sigma_{1}$, which
directly connects to the imaginary part of the dielectric function.
The former quantity shows pronounced resonance at the frequency value
of 180 meV. As we shall see further ahead, these two features can
be retrieved from the polarizability \cite{wang_exciton_2006} of
the excitonic states.

\begin{figure}[h]
\centering{}\includegraphics[width=8.5cm]{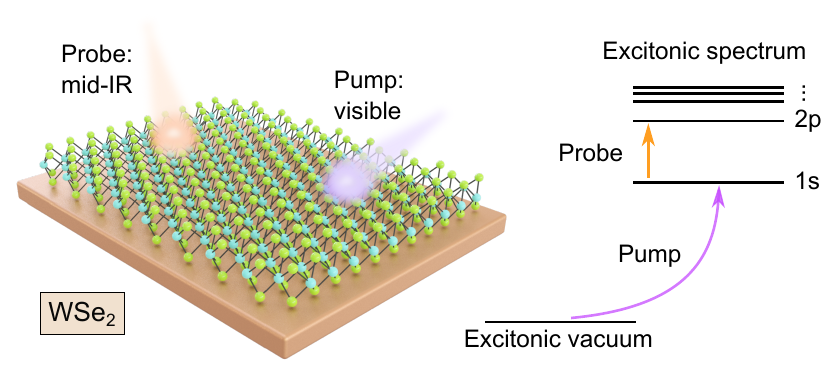}\caption{Pump-probe experiment for observing dark excitonic states in a transition
metal dichalcogenide. The high intensity pump populates the $1$s
ground state of the excitonic energy series, whereas the mid-IR low
intensity probe induces transitions between the excitonic ground state
($1$s) and dark $n$p states, thus allowing to access this usually
not accessible piece the of the dielectric response of the excitonic
gas. This procedure is represented by the energy levels on the right
part of the figure. In the left part of the figure, a sample of WSe\protect\textsubscript{2}
is depicted on which both the pump and the probe laser impinge.}
\end{figure}

The position of excitonic energy levels are known to depend on whether
the system is confined. For example, it is well known that the Rydberg
series of the two-dimensional Hydrogen atom, which in the infinite
systems is given by $E_{n}=-1/[2(n-1/2)^{2}]$ with $n=1,2,3\ldots$,
in atomic units, is strongly affected by confining the atom in a disk
of radius $R$ \cite{chaoscador_two-dimensional_2005}. Recently,
fabrication of transition metal dichalcogenide metamaterials with
atomic precision \cite{munkhbat_transition_2020} has opened the
possibility of studying the dependence of the excitonic energy series
of TMDs upon geometric confinement of different geometries. It is
expected that in addition to negative energy bound states, positive
ones will also appear. As we show ahead, our procedure may be applied
to the study of excitons in circular boxes (which can actually be
obtained, approximately, even in a hexagonal lattice).

In this paper we establish the connection between the polarizability
of an exciton (a microscopic quantity) and the dielectric function,
which is a macroscopic quantity characterizing many particles, when
many excitons, considered non-interacting (and therefore, low density)
form an excitonic gas. We show that our microscopic theory can account
well for the experimental results of Poellmann \emph{et al. }\cite{poellmann_resonant_2015}.
The paper is organized has follows: in section II we present the Fowler's
and Karplus' method which will allow us to compute the optical polarizability
without evaluating a sum over states (whose direct approach is doomed
to fail). This method requires the solution of a differential equation,
which is then used to compute the matrix elements that define the
polarizability. Afterwards, we apply this technique to the simple
case of the cylindrical potential well. The goal of this section is
to benchmark the method. In section III we explore the problem of
excitons in WSe\textsubscript{2}, presenting a way of solving the
Fowler's and Karplus' differential equation for excitonic problems,
computing the polarizability and, from it, the dielectric function
and conductivity of a macroscopic collection of excitons. This section
ends with a comparison of our theoretical results with experimental
data and a good agreement is found. In section IV we give our final
remarks. An appendix showing how to solve the Wannier equation in
log-grid closes the paper.

\section{Karplus' and Fowler's method}

In this section we will briefly present the Karplus' and Fowler's
\cite{Podolsky1928,Karplus1962,karplus_variationperturbation_1963,fowler_energy_1984}
method to compute the dynamical polarizability of different systems.
After the essential features of this approach are laid out, we will
explore the problem of the infinite cylindrical well. This serves
two purposes: on the one hand it allows us to apply the formalism
to a simple yet instructive case; and on the other, it sets the stage
for the problem of two-dimensional excitons that will be treated in
the subsequent section and involves the solution of the Wannier equation.

\subsection{Theory}

Let us begin our discussion considering the following Hamiltonian
written in atomic units (a.u.; this system of units is used throughout
the paper, except when stated otherwise) in the dipole approximation:
\begin{align}
H & =H_{0}-\mathbf{r}\cdot\mathbf{F}(t),\\
 & =-\frac{1}{2\mu}\nabla^{2}+V(\mathbf{r})-\mathbf{r}\cdot\mathbf{F}(t)\label{eq:general H}
\end{align}
where $\mu$ is a mass term, $\nabla^{2}$ is the Laplacian, $V(\mathbf{r})$
is a potential energy term and $\mathbf{F}(t)$ is an external time
dependent harmonic electric field:
\begin{equation}
\mathbf{F}(t)=F\left[e^{-i\omega t}+e^{i\omega t}\right],\label{eq:External_Field}
\end{equation}
with $\omega$ the field's frequency and $F$ its amplitude. This
general Hamiltonian can be used to describe the interaction of different
systems with an external electric field. Two examples which will be
treated in this work are the interaction of an electric field with
a particle trapped inside a cylindrical well, and the interaction
of light with excitons in 2D materials. The first will be treated
ahead in the present section as an introductory problem, while the
latter is the main subject of this work and will be studied in the
following section.

Now, for simplicity, we consider the electric field to be applied
along the $-x$ direction. With this assumption, the time dependent
Schrödinger equation reads:
\begin{equation}
\left[H_{0}+xF(t)\right]|\psi(t)\rangle=i\frac{\partial}{\partial t}|\psi(t)\rangle,\label{eq:TDSE}
\end{equation}
where $|\psi(t)\rangle$ is the state vector describing the wave function
of the system in the presence of the electric field. Following Fowler's
and Karplus' approach, we expand the state vector in powers of $F$
as:
\begin{align}
|\psi(t)\rangle & =e^{-iE_{0}t}|\psi_{0}\rangle\nonumber \\
 & +Fe^{-i(E_{0}-\omega)t}|\psi_{1}^{+}\rangle+Fe^{-i(E_{0}+\omega)t}|\psi_{1}^{-}\rangle+\mathcal{O}\left(F^{2}\right)
\end{align}
where $E_{0}$ and $|\psi_{0}\rangle$ are the energy and state vector,
respectively, of the unperturbed system and $|\psi_{1}^{\pm}\rangle$
are the first order corrections to the state vector. Inserting this
expansion in Eq. (\ref{eq:TDSE}), and keeping only terms up to first
order in $F$, we find:
\begin{align}
H_{0}|\psi_{0}\rangle & =E_{0}|\psi_{0}\rangle\\
H_{0}|\psi_{1}^{\pm}\rangle+x|\psi_{0}\rangle & =\left(E_{0}\mp\omega\right)|\psi_{1}^{\pm}\rangle,\label{eq:psi_1 diff eq}
\end{align}
where the first equation is nothing more than the Schrödinger equation
of the unperturbed system, and the second equation defines the first
order correction $|\psi_{1}^{\pm}\rangle.$ The dynamical polarizability
$\alpha(\omega)$ is defined as \cite{karplus_variationperturbation_1963}:
\begin{equation}
\alpha(\omega)=-\left(\langle\psi_{0}|x|\psi_{1}^{+}\rangle+\langle\psi_{0}|x|\psi_{1}^{-}\rangle\right),\label{eq:Fowler alpha}
\end{equation}
such that:
\begin{equation}
E(\omega)=E_{0}-\frac{1}{2}\alpha(\omega)F^{2}.
\end{equation}
We note that no term proportional to $F$ appears in systems with
inversion symmetry (like the ones we will be considering). Note that
we have departed here from the traditional Rayleigh-Schrödinger perturbation
theory, which goes a step further and expands $\vert\psi_{1}^{\pm}\rangle$
in the basis of the unperturbed Hamiltonian.

As just said, if Rayleigh-Schrödinger time-dependent perturbation
theory had been used to compute the dynamical polarizability, the
final result would consist in a sum over states which, in principle,
could not be exactly computed, since it would require the computation
of an arbitrary large number of matrix elements whose form is unknown
analytically. Also, a numerical approach to the sum over states would
fail, because of the difficulty of describing the wave functions of
the continuum of states. Usually, these sums are truncated and only
the first terms are considered. The quality of this approximation
depends on how fast the sum over states converges to a given value
\cite{Henriques_phosphorene}. In comparison, the presented approach
bypasses the sum over states, and requires the computation of only
two matrix elements in order to obtain the dynamical polarizability,
as described by Eq. (\ref{eq:Fowler alpha}), and, at the same time,
accomplishes the summation over all states.

\subsection{The infinite cylindrical well}

In order to test Eq. (\ref{eq:Fowler alpha}), we will now explore
a simple example: a particle trapped inside a 2D cylindrical potential
well. For simplicity let us consider the particle to be an electron,
such that $\mu=1$ (in atomic units) in Eq. (\ref{eq:general H}).
The potential term reads:
\begin{equation}
V(r)=\begin{cases}
0, & r<R\\
\infty, & r>R
\end{cases},
\end{equation}
where $R$ is the radius of the cylindrical potential well.

Before the dynamical polarizability can be computed, the unperturbed
system has to be studied. Although this is a simple system with a
well established solution, we will give, for completeness, a brief
description of the necessary steps to find the wave functions, and
the respective energies, of the cylindrical potential well. First,
the Schrödinger equation is written in cylindrical coordinates $(r,\theta)$,
and a solution by separation of variables is proposed. After this
is done, two equations arise. The angular equation is satisfied by
a complex exponential $e^{im\theta},$ where $m$ is the angular quantum
number. The radial equation yields Bessel functions of the first kind.
After these two solutions are combined, the boundary condition is
imposed, that is, the requirement that the wave function must vanish
when $r=R$. This condition, similarly to what happens in the 1D infinite
well, defines the energy spectrum. In summary, the wave functions
are:
\begin{equation}
\psi_{nm}(r,\theta)=\frac{\mathcal{C}_{nm}}{\sqrt{2\pi}}J_{m}\left(\frac{q_{mn}r}{R}\right)e^{im\theta},
\end{equation}
where $J_{m}$ is a Bessel function of the first kind, $q_{mn}$ is
the $n-$th zero of $J_{m}(x)$ and $\mathcal{C}_{nm}=\sqrt{2}\left[\left|J_{m+1}(q_{mn})\right|R\right]^{-1}$
is a normalization constant. The energy spectrum reads:
\begin{equation}
E_{nm}=\frac{1}{2}\left(\frac{q_{mn}}{R}\right)^{2}.
\end{equation}
In Table \ref{tab:Cylindrical en} we present the numerical value
of a few energy levels for a system with $R=2$ a.u..

\begin{table}
\centering{}%
\begin{tabular}{cccccc}
\toprule 
\multicolumn{3}{c}{$m=0$} & \multicolumn{3}{c}{$m=1$}\tabularnewline
\midrule
\midrule 
$n=1$ & $n=2$ & $n=3$ & $n=1$ & $n=2$ & $n=3$\tabularnewline
\midrule 
0.723 & 3.809 & 9.361 & 1.835 & 6.152 & 12.937\tabularnewline
\bottomrule
\end{tabular}\caption{\label{tab:Cylindrical en}Numerical values of the energies (in atomic
units) of the infinite cylindrical potential well, for the first three
states with $m=0$ and $m=1$. Transitions from the ground state $(n=1$,
$m=0)$ to the states with $m=1$ require the energies: 1.112, 5.429
and 12.214.}
\end{table}

\begin{figure}[h]
\begin{raggedright}
\includegraphics[width=8.5cm]{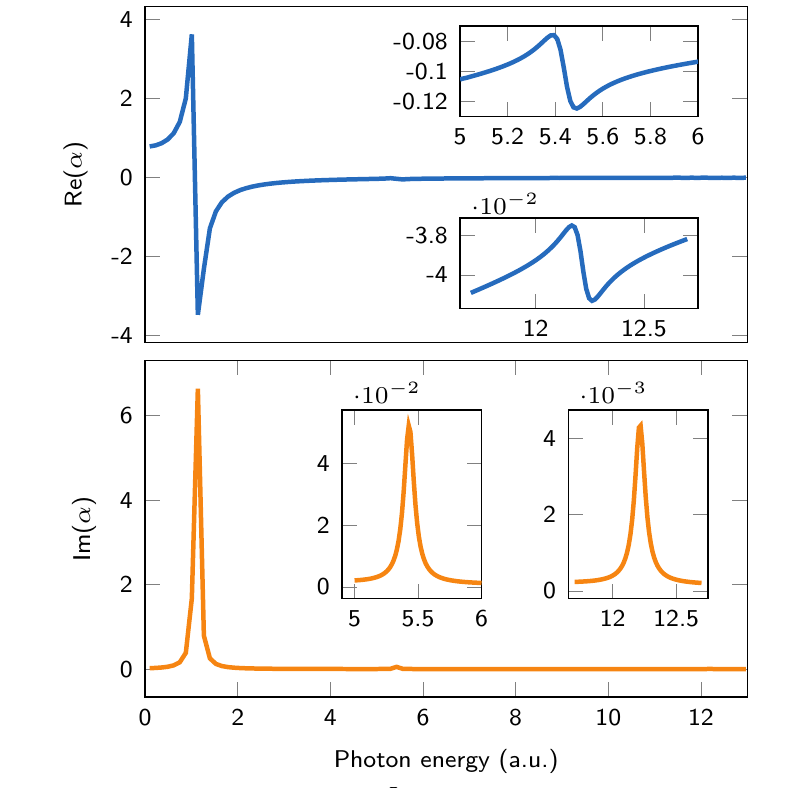}\caption{\label{fig:Cylindrical_Well_Pol}Plot of the real (top) and imaginary
(bottom) parts of the dynamical polarizability for an infinite cylindrical
well, obtained using Eq. (\ref{eq:Fowler alpha}), and with $\psi_{1}^{\pm}$
computed numerically. Both plots present three resonances whose positions
exactly coincide with the energies associated with transitions from
the state with $m=0$ and $n=1$, to the three states with $m=1$
presented in Table \ref{tab:Cylindrical en}. In order to obtain the
real and imaginary parts of the polarizability, a small imaginary
term, $\delta=0.05$ a.u., was subtracted to $\omega$. We considered
the particle to be an electron, confined to a disk of radius $R=2$
a.u..}
\par\end{raggedright}
\end{figure}

Now that we are in possession of the unperturbed wave functions and
energy spectrum, the computation of the dynamical polarizability due
to the presence of a time dependent external field can be performed.
In agreement with the previous section, we will consider a harmonic
electric field aligned along the $x-$direction. Furthermore, we will
be concerned with the polarizability of the ground state, which in
this case corresponds to the state with $m=0$ and $n=1$. To compute
$\alpha(\omega)$ we will have to compute the matrix elements $\langle\psi_{10}|x|\psi_{1}^{\pm}\rangle$.
To do so we must first solve Eq. (\ref{eq:psi_1 diff eq}) and find
$|\psi_{1}^{\pm}\rangle$. For $0<r<R$ we have:
\begin{equation}
\left(\frac{\mathbf{p}^{2}}{2}-E_{10}\pm\omega\right)\psi_{1}^{\pm}(\mathbf{r})=-r\cos\theta\psi_{10}(\mathbf{r}).
\end{equation}
We now set $\psi_{1}^{\pm}(\mathbf{r})$ as $\left[\mathcal{P_{\pm}}(r)/\sqrt{r}\right]\cos\theta$.
This transformation has two advantages, as it takes care of the angular
part of the differential equation, while simultaneously eliminating
any term proportional to a first derivative in $r$. With this assumption
we find:
\begin{equation}
8r^{7/2}\psi_{10}(r)+\left[3+8r^{2}(-E_{0}\pm\omega)\right]\mathcal{P}_{\pm}(r)=4r^{2}\mathcal{P_{\pm}}''(r),\label{eq:cyl_well diff eq}
\end{equation}
subject to the boundary conditions $\mathcal{P}(0)=0$ and $\mathcal{P}(R)=0$.
A numerical solution to this differential equation can be easily obtained,
and once this is done, computing the polarizability is a trivial step,
amounting to a numerical quadrature. In Fig. \ref{fig:Cylindrical_Well_Pol}
we plot the polarizability obtained after the differential equation
was solved numerically. In order to obtain the real and imaginary
parts of the polarizability, a small imaginary term was subtracted
to $\omega$, that is $\omega\rightarrow\omega-i\delta$, with $\delta\ll\omega$;
this parameter controls the linewidth of the resonances, and should
be chosen to match experimental results. Since our goal is not to
compare this simple example with experimental data we simply set $\delta=0.05$
a.u.. Inspecting the results in Fig. \ref{fig:Cylindrical_Well_Pol},
we find that, as expected, clear resonances appear at energies corresponding
to transitions from the ground state to states with $m=1$, as can
be confirmed by recalling the values presented in Table \ref{tab:Cylindrical en}.
The first resonance, corresponding to a transition from the ground
state to the state with $n=m=1$ is by far the dominant one, being
orders of magnitude more intense than the second and third peaks.
Although not depicted, at higher energies other peaks with even smaller
oscillator strengths appear.

\section{Application to two-dimensional excitons}

In the present section we will apply the same theoretical ideas developed
so far to compute the optical polarizability of excitons in 2D materials.
Contrary to the case of the infinite cylindrical potential well, the
existence of a Coulomb-like potential combined with an infinite geometry
transforms this problem into a rather involved one. To cope with the
inherent difficulties, we will consider the problem to be confined
on a finite disk, and solve the Fowler-Karplus' differential equation
using a variational approach. In the end we will consider the radius
of the disk to be sufficiently large so that the obtained results
are numerically identical to the ones we would obtain with an infinite
geometry, allowing us to compare our predictions with the experimental
results of Poellmann \emph{et al} \cite{poellmann_resonant_2015}.

\subsection{Relation between polarizability, dielectric function, and optical
conductivity }

Consider a gas with $N_{X}$ excitons per unit area in a two dimensional
material. The concentration $N_{X}$ is controlled by the interband
excitation process (fluence of the laser), the exciton recombination,
and disorder. If the electric field creating the exciton gas is weak,
the excitons' concentration is small and the gas can be considered
non interacting. The polarization of a 2D material due to \emph{intra-exciton
transitions} is defined as
\begin{equation}
\mathbf{P}_{{\rm ex}}^{{\rm intr.}}=N_{X}\mathbf{p}=N_{X}\alpha\mathbf{E},
\end{equation}
where $\mathbf{p}$ is the dipole moment of the exciton, in units
of charge times length and $\alpha$ is the polarizability, in unites
of $\epsilon_{0}$ times volume, with $\epsilon_{0}$ the dielectric
vacuum permittivity. The susceptibility of the exciton gas associated
with intra-excitons transitions is defined has
\begin{equation}
\chi_{{\rm ex}}^{{\rm intra.}}=\frac{\mathbf{P}_{{\rm ex}}^{{\rm intr.}}}{\epsilon_{0}\mathbf{E}}=N_{X}\alpha/\epsilon_{0},
\end{equation}
and has units of length. We note that, in addition to $\chi_{{\rm ex}}^{{\rm intra.}}$,
there are other terms contributing to the total susceptibility of
the excitons gas: transition from the exciton vacuum to the $n$s
levels and to other even parity states, and the background susceptibility,
accounting for transitions to higher-energy bands, as well as other
processes contributing to the total dielectric function of a TMD \cite{GU2019}.
These, however, are not the focus of this paper and have been addressed
elsewhere \cite{Itai2020}. Thus, we introduce the dielectric function
associated with intra-exciton transition in a diluted medium (that
is, low exciton density) as,
\begin{equation}
\Delta\epsilon(\omega)=\chi_{{\rm ex}}^{{\rm intra.}}/d=\alpha\frac{N_{X}}{\epsilon_{0}d}\frac{1}{g_{s}g_{v}},\label{eq:pol_to_eps}
\end{equation}
where $d$ is the effective thickness of the TMD layer and $g_{s}$
and $g_{v}$ are the spin and valley degeneracy factors.The real and
imaginary parts of the dielectric function due to intra-exciton transitions
follow from this equation. The relation between the dielectric function
and the conductivity, $\sigma$, follows from \cite{Kaind2009} $\Delta\epsilon=i\sigma/(\epsilon_{0}\omega)$,
and we obtain
\begin{equation}
\sigma=-i\epsilon_{0}\omega\Delta\epsilon.\label{eq:eps_to_cond}
\end{equation}
The computation of these two quantities, $\Delta\epsilon(\omega)$
and $\sigma$, for the $1$s$\rightarrow n$p excitonic transitions
is our main goal. (It would be interesting to derive a Clausius-Mossotti
\cite{ryazanov_clausius-mossotti-type_2006}, relation for two-dimensional
excitons, although this is not the focus of our work.)

\subsection{The variational wave function to the ground state of the exciton}

In what follows we will compute the optical polarizability of the
excitonic ground state ($1$s state). In order to do so, we need to
first determine its wave function. Finding an analytical expression
for the $1s$ exciton state is not a simple task \cite{henriques_optical_2019,Henriques2020},
and because of that we will consider the following double exponential
variational \textit{ansatz} to describe the ground state \cite{pedersen_exciton_2016}
(see also \cite{quintela_colloquium_2020}):
\begin{equation}
\psi_{0}=\mathcal{N}\left(e^{-ar}-be^{-\gamma ar}\right),
\end{equation}
where $\mathcal{N}$ is a normalization constant and $a,$ $b$, and
$\gamma$ are variational parameters obtained from energy minimization.
This variational \textit{ansatz} is considerably more accurate than
if a single exponential is used. In the context of the excitonic problem,
the Hamiltonian given in Eq. (\ref{eq:general H}) becomes:
\begin{equation}
H=-\frac{1}{2\mu}\nabla^{2}+V_{{\rm RK}}(r)-\mathbf{r}\cdot\mathbf{F}(t),
\end{equation}
where $\mu$ is the reduced mass of the electron-hole system, $\mathbf{F}(t)$
is the external field previously defined in Eq. (\ref{eq:External_Field}),
and $V_{{\rm RK}}(r)$ is the Rytova-Keldysh potential \cite{rytova1967,keldysh1979coulomb}:
\begin{equation}
V_{{\rm RK}}(r)=-\frac{\pi}{2r_{0}}\left[\mathbf{H}_{0}\left(\frac{\kappa r}{r_{0}}\right)-Y_{0}\left(\frac{\kappa r}{r_{0}}\right)\right],\label{eq:RK-potential}
\end{equation}
with $\kappa$ the mean dielectric constant of the media above and
below the TMD, $r_{0}$ is an intrinsic parameter of the 2D material
which can be related to the effective layer thickness $d$ \cite{poellmann_resonant_2015},
and $\mathbf{H}_{0}$ and $Y_{0}$ are the Struve $H$-function of
zero-th order and the Bessel function of the second kind of the same
order, respectively. This potential is the solution of the Poisson
equation for a charge embedded in a thin film. For large distances
the Rytova-Keldysh potential presents a Coulomb tail, and diverges
logarithmically near the origin.

Having determined a way to compute the excitonic ground state wave
function and binding energy, we turn our attention to solving the
Fowler's and Karplus' \cite{fowler_energy_1984,karplus_variationperturbation_1963}
differential equation introduced in Eq. (\ref{eq:psi_1 diff eq}).
Comparing the excitonic problem with the simple example treated in
the previous section, we stress two key differences. While in the
case of the cylindrical well our problem had a bounded geometry, that
is, the domain of our problem was necessarily restricted to $r<R$,
in the present scenario we find an unbounded problem, whose domain
extends up to infinity. Moreover, in the previous section no potential
term appeared in our calculations, while in the present case we find
the Rytova-Keldysh potential which, as previously noted, presents
a Coulomb$-1/r$ tail at large distance and a $\ln(r)$ divergence
at short distances. This slow decaying behavior is known to be problematic
in numerical calculations. Considering these two remarks, it is clear
that finding the solution to the differential equation will not be
as straight forward a process as it was before.

The first step we take in order to proceed with the calculations is
to confine our system on a finite disk with large radius $R$. In
practice, when numerical values are introduced, we will consider $R$
to be sufficiently large in order to obtain results which are numerically
identical to the ones that one would find for an unbounded problem,
and thus be able to compare our results with those from Ref. \cite{poellmann_resonant_2015}.
This approach would also prove its usefulness if a genuinely finite
system was studied. The confinement is reflected on the wave function
of the ground state by the introduction of an additional multiplicative
term $(r-R)$, that is:
\begin{equation}
\psi_{0}^{{\rm fin.}}(r)=\mathcal{N}^{{\rm fin}}\left(e^{-ar}-be^{-\gamma ar}\right)\left(r-R\right).
\end{equation}
For a sufficiently large $R$ one finds:
\begin{equation}
\lim_{R\rightarrow\infty}\psi_{0}^{{\rm fin.}}(r)=\psi_{0}(r).
\end{equation}
As before, the parameters $a,$ $b$ and $\gamma$ are determined
from energy minimization. This transformation does not allow us to
immediately solve the differential equation. However, it unlocks a
new way of approaching the problem.

\subsection{The dynamic variational method}

Since we are now effectively working on a finite disk, and we saw
in the previous section that Bessel functions are an appropriate complete
set of functions to describe a problem in such a geometry, we propose
that $\psi_{1}^{\pm}(\mathbf{r})$ can be written as:
\begin{equation}
\psi_{1}^{\pm}(\mathbf{r})=\cos\theta\sum_{n=1}^{N}c_{n}^{\pm}J_{1}\left(\frac{z_{1n}r}{R}\right),\label{eq:psi_1 J_1}
\end{equation}
where $J_{1}(z)$ is a Bessel function of the first kind, $z_{1n}$
is the $n-$th zero of $J_{1}(z)$, $R$ is the radius of the disk,
$N$ is the number of Bessel function we choose to use, and $\left\{ c_{n}^{\pm}\right\} $
are a set of coefficients yet to be determined. As proposed in Ref.
\cite{karplus_variationperturbation_1963,montgomery_one-electron_1978,yaris_timedependent_1963,yaris_timedependent_1964}
the values of $\left\{ c_{n}^{\pm}\right\} $ are determined from
the minimization of the following functional:
\begin{align}
\mathcal{J}_{\pm} & =\int d\mathbf{r}\psi_{1}^{\pm}(\mathbf{r})\left[H_{0}-E_{0}\pm\hbar\omega\right]\psi_{1}^{\pm}(\mathbf{r})\nonumber \\
 & +2\int d\mathbf{r}\psi_{1}^{\pm}(\mathbf{r})r\cos\theta\psi_{0}(\mathbf{r}).
\end{align}
We now recall the following relation regarding the orthogonality of
Bessel functions on a finite disk of radius $R$ \cite{zwillinger_table_2014}:
\begin{equation}
\int_{0}^{R}J_{\nu}\left(\frac{\alpha_{\nu m}r}{R}\right)J_{\nu}\left(\frac{\alpha_{\nu n}r}{R}\right)rdr=\frac{R^{2}}{2}\delta_{nm}\left[J_{\nu+1}(\alpha_{\nu m})\right]^{2},
\end{equation}
where $\alpha_{\nu m}$ is the $m-$th zero of $J_{\nu}(z).$ Using
this relation, one easily shows that the functional can be rewritten
as:
\begin{align}
\mathcal{J}_{\pm} & =\frac{\pi R^{2}}{2}\sum_{n=1}\left(c_{n}^{\pm}\right)^{2}\left[\frac{\hbar^{2}}{2\mu}\frac{z_{1n}^{2}}{R^{2}}-E_{0}\pm\hbar\omega\right]\left[J_{2}\left(z_{1n}\right)\right]^{2}\nonumber \\
 & +\pi\sum_{n=1}\sum_{k=1}c_{n}^{\pm}c_{k}^{\pm}\mathcal{I}_{kn}+2\pi\sum_{n=1}c_{n}^{\pm}\mathcal{S}_{n},
\end{align}
where $\mathcal{S}_{n}$ and $\mathcal{I}_{kn}$ refer to the following
integrals involving one and two Bessel functions, respectively:
\begin{align}
\mathcal{I}_{kn} & =\int J_{1}\left(\frac{z_{1k}r}{R}\right)V_{{\rm RK}}(r)J_{1}\left(\frac{z_{1n}r}{R}\right)rdr\label{eq: I integral}\\
\mathcal{S}_{n} & =\int J_{1}\left(\frac{z_{1n}r}{R}\right)\psi_{0}^{{\rm fin.}}(r)r^{2}dr.\label{eq:S integral}
\end{align}
Remembering our goal of minimizing the functional $\mathcal{J}_{\pm}$,
we must now differentiate it with respect to the different $c_{n}^{\pm}$
with $n\in\left\{ 1,2,...,N\right\} $. Doing so, and noting in passing
that $\mathcal{I}_{kn}=\mathcal{I}_{nk}$, one finds:
\begin{align}
c_{j}^{\pm}\left\{ \frac{R^{2}}{2}\left[\frac{1}{2\mu}\frac{z_{1j}^{2}}{R^{2}}-E_{0}\pm\omega\right]\left[J_{2}\left(z_{1j}\right)\right]^{2}+\mathcal{I}_{jj}\right\} \nonumber \\
+\sum_{n\neq j}^{N}c_{n}^{\pm}\mathcal{I}_{jn}=-\mathcal{S}_{j}
\end{align}
with $j\in\left\{ 1,2,...,N\right\} $. It is now clear that this
defines a linear system of equations whose solution determines the
values of the coefficients $c_{n}^{\pm}$. Moreover, we can write
this concisely using matrix notation, as follows:
\begin{equation}
\mathbb{M}\cdot\mathbf{c}^{\pm}=-\mathbf{S},
\end{equation}
where $\mathbf{c}^{\pm}$ and $\mathbf{S}$ are column vectors defined
as:
\begin{align}
\left[\mathbf{c}^{\pm}\right]^{{\rm T}} & =\left(c_{1}^{\pm},c_{2}^{\pm},\ldots,c_{N}^{\pm}\right)\\
\mathbf{S}^{{\rm T}} & =\left(\mathcal{S}_{1},\mathcal{S}_{2},\ldots,\mathcal{S}_{N}\right),
\end{align}
and $\mathbb{M}$ is an $N\times N$ matrix with:
\begin{equation}
\left(\mathbb{M}\right)_{ij}=\begin{cases}
g_{i}^{\pm}(\omega)+\mathcal{I}_{ii} & i=j\\
\mathcal{I}_{ij} & i\neq j
\end{cases},
\end{equation}
where $g_{i}^{\pm}(\omega)=R^{2}\left[z_{1i}^{2}/(2\mu R^{2})-E_{0}\pm\omega\right]/2$.
After $\mathbb{M}$ and $\mathbf{S}$ are computed using Eqs. (\ref{eq: I integral})
and (\ref{eq:S integral}), respectively, the coefficients that determine
$\psi_{1}^{\pm}(\mathbf{r})$ are readily obtained:
\begin{equation}
\mathbf{c}^{\pm}=-\mathbb{M}^{-1}\cdot\mathbf{S},
\end{equation}
and the solution of the differential equation is found. As was previously
mentioned, when specific values are introduced in the problem, the
radius $R$ must be chosen large enough in order to produce results
which are identical to the ones of an unbounded system. As the value
of $R$ increases, the value of $N$ must also increase in order to
accurately describe the desired physical system. At first, this may
seem problematic, since for each value of $\omega$ we would have
to compute $\mathbb{M}$ (with $N^{2}$ entries) and solve the linear
system associated with it. However, two aspects prevent this approach
of becoming inadequately inefficient as $N$ increases: i) the matrix
$\mathbb{M}$ is symmetric, reducing the number of independent entries
from $N^{2}$ to $N(N+1)/2$; ii) more importantly, the most time
consuming part of the calculation, computing all the relevant $\mathcal{I}_{ij}$,
is independent of $\omega$, and does not need to be reevaluated every
time a new energy is used.

\begin{table}[h]
\begin{centering}
\begin{tabular}{ccccccc}
\toprule 
$\mu/m_{0}$ & $r_{0}$ /$a_{0}$ & $\kappa$ & $N_{X}$ (cm$^{-2}$) & $d/a_{0}$ & $R$ /$a_{0}$ & $N$\tabularnewline
\midrule
\midrule 
0.167 & $54$ & 3.32 & 1.8$\times10^{12}$ & 6.04 & 800 & 35\tabularnewline
\bottomrule
\end{tabular}
\par\end{centering}
\caption{\label{tab:Parameters for WSe2}Parameters used to describe excitons
in WSe\protect\textsubscript{2}. The value of $R$ was chosen in
such a way that the wave functions of the first excited states are
effectively zero even before $r=R$. Inspecting Figure \ref{fig: Shooting WF},
we observe that the states with $m=1$ and $n=2,3$ have wave functions
which essentially vanish for $r\sim400a_{0}$, where $a_{0}\sim0.53$
$\AA$ is the Bohr radius. The number of Bessel functions ($N$)
was chosen as the minimum number which upon addition of more functions
leaves the final result unchanged. The parameters related with WSe\protect\textsubscript{2}were
taken from Ref. \cite{poellmann_resonant_2015}. The reduced mass
$\mu$ is given in units of the bare electron mass $m_{0}$.}
\end{table}

At last, to compute the dynamical polarizability we look back at Eqs.
(\ref{eq:Fowler alpha}) and (\ref{eq:psi_1 J_1}) and write:
\begin{equation}
\alpha(\omega)=-2\pi\left(\mathbf{c}^{+}(\omega)+\mathbf{c}^{-}(\omega)\right)\cdot\mathbf{S},
\end{equation}
Now, to put together everything we developed so far to the test, let
us consider the specific case of excitons in WSe\textsubscript{2}.
The different parameters that characterize this problem are shown
in Table \ref{tab:Parameters for WSe2}. Note that the value of the
radius was chose as $R=800a_{0},$ with $a_{0}\sim0.53$ $\AA$,
the Bohr radius. The choice of this value is understood from the inspection
of Fig \ref{fig: Shooting WF}, in the Appendix, where we present
numerically obtained wave functions for the first excited states of
WSe\textsubscript{2} (considering an unbounded system). There we
observe that the wave functions are essentially zero for $r\sim400a_{0}$,
and thus the choice of $R=800a_{0}$ renders our problem effectively
infinite. Moreover, in Table \ref{tab:Parameters for WSe2} we also
observe that $N=35$ Bessel functions were used. This value was chosen
since it corresponds to the minimum value of functions upon which
addition of more functions leaves the final result invariant. Inspection
of the values of $\mathbf{c}^{\pm}$ allowed us to conclude that the
method converged.

\subsection{Results}

\begin{figure}[h]
\begin{raggedright}
\includegraphics[width=8.5cm]{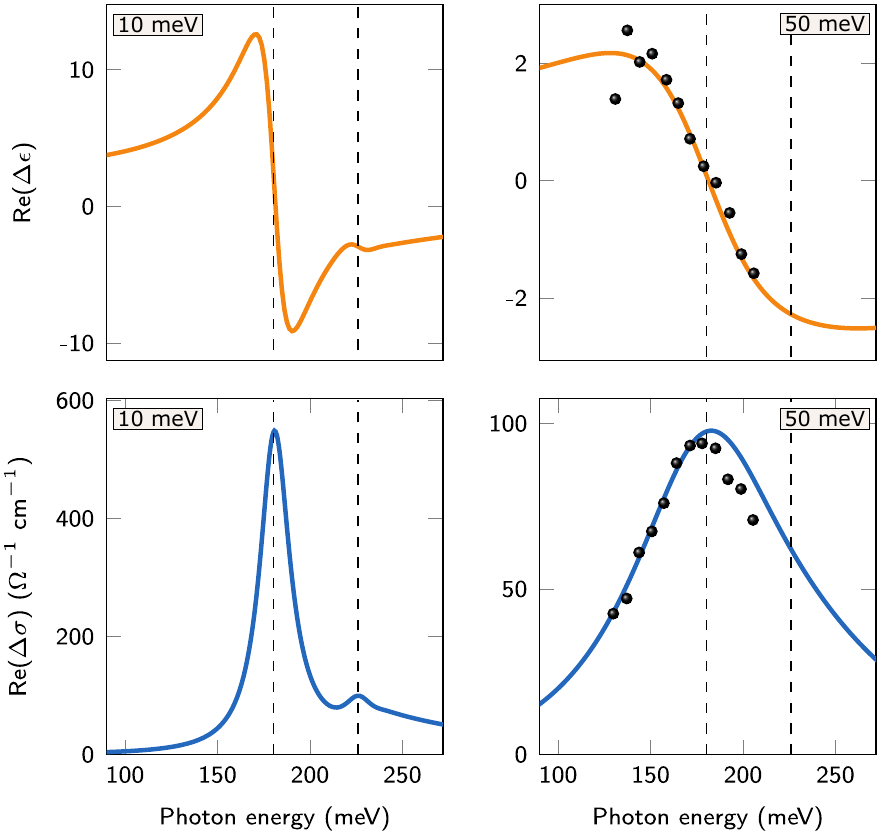}\caption{\label{fig:WSe_2 plot}Real part of the dielectric function (top)
and optical conductivity (bottom) for two different linewidths. 10meV
(left) and 50meV (right). When a broadening of 10 meV is considered
two resonances are clearly visible, appearing at energies corresponding
to transitions from the ground state (1$s$) to the states $2p$ and
$3p$. When a broadening of 50 meV is used, only the main resonance
is resolved. For this case we also depict the experimental values
measured in Ref. \cite{poellmann_resonant_2015} (large black dots).
The agreement between our theoretical prediction and the experimental
points is remarkable. The parameters of Table \ref{tab:Parameters for WSe2}
were used. The energies corresponding to the transitions above mentioned
are given in Table \ref{tab:Binding-energies}. The vertical dashed
lines represent the energy difference $E=E_{n{\rm p}}-E_{1{\rm s}}$
as determined from the solution of the Wannier equation. As shown
in the experiments \cite{poellmann_resonant_2015}, the there is
a relation between the level broadening and the exciton's concentration.
Therefore, the panels on the left-hand side of this figure do not
represent a realistic experimental situation but rather show the role
that reducing the broadening has in revealing higher energy dark states. }
\par\end{raggedright}
\end{figure}

In Fig. \ref{fig:WSe_2 plot} we depict the real part of the dielectric
function and the optical conductivity, obtained from the real and
imaginary parts of the the polarizability using Eqs. (\ref{eq:pol_to_eps})
and (\ref{eq:eps_to_cond}), for two different linewidths, 10 meV
and 50 meV, and compare our theoretical prediction with experimental
results for the latter case. The linewidths are introduced in the
problem by subtracting a small imaginary part (10 or 50 meV in the
present case) to the photon energy. This is also what allows us to
extract both the real and imaginary parts from Eq. (\ref{eq:pol_to_eps}).
To model the problem the values of Table \ref{tab:Parameters for WSe2}
are used.

Let us start by looking at the cases where a broadening of 10 meV
was used. Similarly to what was found for the cylindrical well, we
observe two clearly visible resonances at the energies corresponding
to transitions from the excitonic ground state to the first two excited
states with $m=1$, that is the $2$p and $3$p states. As expected,
the resonance associated with the $1s\rightarrow2p$ transition is
significantly more intense than the others. More resonances exist
after the second one, but are not resolved due to the value of the
broadening parameter and their smaller oscillator strength. No resonances
appear above the excitonic ground state energy. If the value of $R$
was chosen smaller this may not have been the case due to the appearance
of states with positive energy as a consequence of the confinement
\cite{aquino_accurate_2005}, however, these extra resonances would
be barely visible due to the smaller oscillator strengths. The values
of the binding energies of the different excitonic states were computed
numerically using the Numerov-shooting method described in the Appendix.

When $\Delta=50$ meV is used (a value fixed by the phenomenological
fit performed in Ref. \cite{poellmann_resonant_2015}) we observe
a different scenario from the previous one. Due to the increased value
of the linewidth, and the way we introduced this parameter in the
calculations, the $1s\rightarrow2p$ resonance is shifted away from
its theoretical value (computed numerically using the shooting method
described in the Appendix). This artifact needs to be manually corrected.
Contrary to the previous case, this time only the main resonance is
resolved. This is a consequence of the high value of the broadening
parameter considered, as well as the proximity of the different resonances
and their small oscillator strength when compared to the 1s$\rightarrow$2p
transition. Moreover, for this specific scenario, we also depict the
experimental results measured in Ref. \cite{poellmann_resonant_2015},
where the authors found that the resonance is well described by phenomenologically
modeling the $1s\rightarrow2p$ transition with a Lorentzian oscillator.
A good agreement between our theoretical prediction and the experimental
result is visible. Our microscopic theory captures both the position
and magnitude of the measured resonances.

\section{Conclusions}

In this paper we have used Fowler's and Karplus' methods to access
the polarizability of 2D excitons when transitions from bright ($n$s)
states to dark ($n$p) states take place.

For benchmarking this method we first applied it to a single particle
in a two dimensional disk of radius $R$ with vanishing Dirichlet
boundary conditions (we note that the optical properties of condensed
matter systems are sensitive to the boundary conditions \cite{montgomery_frequency-dependent_2016}).
We showed that the peaks in the polarizability are in good agreement
with the energy difference between the ground state and the excited
states of the particle in the disk.

Afterwards, we applied Fowler's and Karplus' method to the central
problem of our work, that of finding the dielectric function of the
exciton gas. To do so, we looked for the transitions from bright ($n$s)
to dark ($n$p) states, which can be accessed in pump-probe experiments.
We chose the transition metal dichalcogenide WSe$_{2}$ as our test
subject, since measurements of the dielectric function due to the
exciton gas have been made recently \cite{poellmann_resonant_2015}.
We found a good agreement between the results obtained from our microscopic
theory and the experimental data. Indeed the Fowler-Karplus approach
allows us to access the excitonic response of the system without much
work, only requiring the knowledge of the ground state of the excitonic
problem. This is a major advantage over the overwhelming fully computational
calculations based on \emph{ab-initio} methods. For making our work
mostly analytical, we use an analytical variational wave function
to describe the ground state of the exciton problem. The numerical
part of the calculation consists on computing and solving a well behaved
linear system of equations. The role of disorder on the visibility
of $1$s$\rightarrow n$p transitions was also discussed. We showed
that for small disorder, $\Delta=10$ meV, several transitions to
different $n$p states are visible in the polarizability spectrum,
whereas for large disorder, $\Delta=50$ meV, the first transition
$1$s$\rightarrow2$p is rather broad, as seen in the experiments,
and higher order transition are masked by the linewidth of the peak.

Finally, we speculate that a similar experiment made on WSe$_{2}$
encapsulated in boron-nitride will allow the revealing of the transition
to higher $n$p rather than just the $1$s$\rightarrow2$p transition.
The nonlinear response of the exciton gas \cite{mossman_dalgarnolewis_2016,kocherzhenko_unraveling_2019}
will be the subject of a future work.

\section*{Acknowledgements}

N.M.R.P acknowledges support by the Portuguese Foundation
for Science and Technology (FCT) in the framework of the Strategic
Funding UIDB/04650/2020.  J.C.G.H.  acknowledges the Center of Physics for a grant
funded by the UIDB/04650/2020 strategic project.
N.M.R.P. acknowledges support
from the European Commission through the project ``Graphene-Driven
Revolutions in ICT and Beyond'' (Ref. No. 881603, CORE 3), COMPETE
2020, PORTUGAL 2020, FEDER and the FCT through projects POCI-01-0145-FEDER-028114 and PTDC/NAN-OPT/29265/2017.

\appendix

\section{Numerov shooting method and the Schrödinger equation in a log-grid}

In this appendix we give a description of the method used to compute
the binding energies of excitons in WSe\textsubscript{2}. To compute
these energies one needs to solve the Wannier equation (in atomic
units):
\begin{equation}
-\frac{1}{2\mu}\nabla^{2}\psi(\mathbf{r})+V_{{\rm RK}}(r)\psi(\mathbf{r})=E\psi(\mathbf{r}),
\end{equation}
where, as in the main text, $\mu$ is the reduced mass of the electron-hole
system, $V_{{\rm RK}}$ is the Rytova-Keldysh potential defined in
Eq. (\ref{eq:RK-potential}), $\psi(\mathbf{r})$ is the exciton wave
function, and $E$ its binding energies.

In order to solve this equation, we propose the usual solution by
separation of variables, such that $\psi(\mathbf{r})=R(r)\Theta(\theta)$.
The angular contribution trivially yields:
\begin{equation}
\Theta(\theta)=\frac{e^{im\theta}}{\sqrt{2\pi}},
\end{equation}
where $m=0,\pm1,\pm2,...$ is the angular quantum number, and the
$\sqrt{2\pi}$ is a normalization factor. Considering the definition
$u(r)=R(r)\sqrt{r}$, the radial equation becomes:
\begin{equation}
-\frac{1}{2\mu}\frac{d^{2}u}{dr^{2}}+\left[\frac{m^{2}}{2\mu r^{2}}-\frac{1}{8\mu r^{2}}+V_{{\rm RK}}(r)\right]u(r)=Eu(r).\label{eq:f diff eq}
\end{equation}
Now, we found it useful to introduce the following change of variable:
\begin{equation}
x=\ln r,
\end{equation}
which effectively transforms our problem from a linear to a logarithmic
one. This transformation proved to be useful in the stabilization
of the numerical calculations due to the divergent behavior of the
Rytova-Keldysh potential at the origin. Moreover, we introduce an
auxiliary function, $f,$ defined as:
\begin{equation}
u(x)=f(x)e^{x/2}.
\end{equation}
Doing so, the radial differential equation is transformed into:
\begin{equation}
-\frac{1}{2\mu}f''(x)+\left[\frac{m^{2}}{2\mu}+e^{2x}V_{{\rm RK}}(e^{x})\right]f(x)=e^{2x}Ef(x),
\end{equation}
with boundary conditions $f(\pm\infty)=0$. This is the equation that,
when solved, defines the binding energies for the excitons in WSe$_{2}$.
This equation performs outstandingly for the usually problematic $m=0$
energy levels, where the centrifugal barrier is absent.

To solve this equation we use the shooting method coupled to a Numerov
algorithm \cite{izaac_computational_2018,berghe_numerical_1989,johnson_renormalized_1978,johnson_new_1977,pillai_matrix_2012}.
Given an initial energy guess, we integrate Eq. (\ref{eq:f diff eq})
from left to right, and from right to left, and match the logarithmic
derivatives of two solutions somewhere sufficiently away from the
edges. The matched wave function will only represent a true bound
state, with a correct binding energy, when both the wave function
itself as well as its derivative are continuous across the whole domain.
If this condition is not satisfied, a new energy guess must be used
and the process repeated. The choice of the starting points of integration
is of crucial importance to obtain faithful results. The staring point
on the left should be chosen as negative as possible in order to accurately
capture the behavior near $r=0$ (or $x\rightarrow-\infty$). In practice,
values such as $x_{{\rm min}}=-10$ already produce excellent results.
The starting point on the right should be chosen such that the wave
function has effectively reached zero significantly before $x_{{\rm max}}.$
Using semi-analytical methods, such as the one studied in Ref. \cite{henriques_optical_2019},
allows one to compute a first guess of the wave function, and thus
determine what value of $x_{{\rm max}}$ should be chosen in the application
of the shooting-Numerov method. The method of Ref. \cite{henriques_optical_2019}
is also a good starting point for the initial guesses of the binding
energies. In Ref. \cite{izaac_computational_2018} an efficient algorithm
to find the binding energies of Coulomb-like problems is described.
The method is numerically stable and fast.

\begin{figure}[h]
\raggedright{}\includegraphics[width=8.5cm]{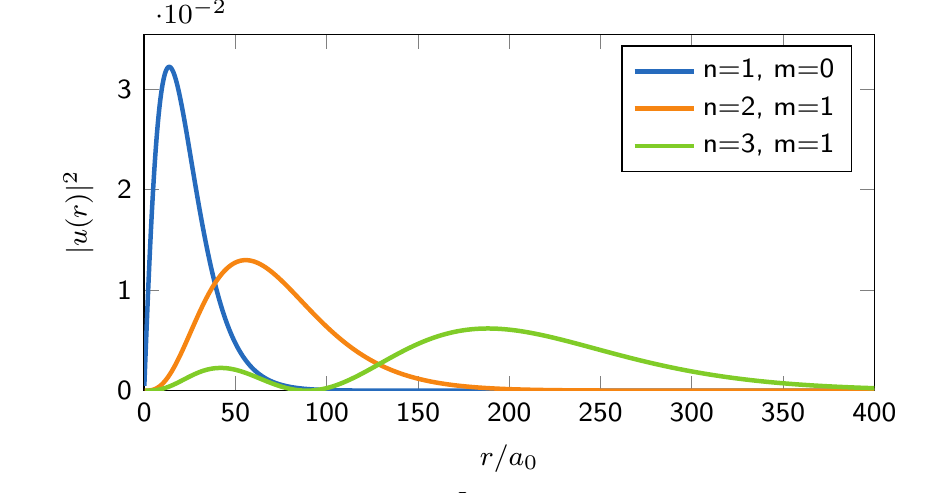}\caption{\label{fig: Shooting WF}Probability radial density of the the ground
state ($n=1$ and $m=0$) and the first excited states with $m=1$,
of excitons in WSe\protect\textsubscript{2}, obtained using the shooting
method. The parameters of Table \ref{tab:Parameters for WSe2} were
used. The radius is given in units of Bohr's radius $a_{0}\approx0.53$
$\AA$.}
\end{figure}

Using the previously described method, and the parameters presented
in Table \ref{tab:Parameters for WSe2} of the main text, we were
able to compute the wave functions of the ground state, and the first
two excited states with $m=1$, of excitons in WSe\textsubscript{2}
depicted in Figure \ref{fig: Shooting WF}. As expected, these wave
functions resemble those one would obtain with the Coulomb potential.
However, since the Rytova-Keldysh potential originates states with
smaller binding energies than the Coulomb potential, these wave functions
are more spread out in space. Although only three wave functions are
presented, this method also allows the computation of higher excited
states.

\begin{table}[h]
\centering{}%
\begin{tabular}{ccccc}
\toprule 
$m=0$ & \multicolumn{4}{c}{$m=1$}\tabularnewline
\midrule
\midrule 
$n=1$ & $n=2$ & $n=3$ & $n=4$ & $n=5$\tabularnewline
\midrule 
$-255$ & $-75$ & $-29$ & $-15$ & $-9$\tabularnewline
\bottomrule
\end{tabular}\caption{\label{tab:Binding-energies}Binding energies (in meV) of a few excitonic
states in WSe\protect\textsubscript{2}, obtained with the shooting
method. The parameters of Table \ref{tab:Parameters for WSe2} were
used. The energy difference $\Delta E=-75+255=180$ meV coincides
very well with the value observed in the experiment at moderate densities
of the exciton gas.}
\end{table}

The binding energies of the ground state, as well as the first states
with $m=1$, are presented in Table \ref{tab:Binding-energies}. These
energies, and in particular those corresponding to the transitions
$1s\rightarrow2p$ and $1s\rightarrow3p$, are in agreement with Figure
\ref{fig:WSe_2 plot} of the main text.




%

\end{document}